\documentclass{iopart}

\usepackage{graphicx}

\usepackage{iopams}

\newcommand{\RM}{{\mathbb R}}
\newcommand{\ZM}{{\mathbb Z}}
\newcommand{\QM}{{\mathbb Q}}
\newcommand{\CM}{{\mathbb C}}
\newcommand{\TM}{{\mathbb T}}

\newcommand{\NMp}{{{\mathbb N}_+}}
\newcommand{\LM}{{\mathbb L}}
\newcommand{\TMN}{{{\mathbb T}^{n-1}}}

\newcommand{\dd}{{\rm d}}
\newcommand{\ddb}{{\rm {\bf d}}}
\newcommand{\vt }{\vec \theta}
\newcommand{\ba}{{\bar \alpha}}
\newcommand{\vz}{{\varphi}}
\newcommand{\vj}{{\varphi _j}}
\newcommand{\vl}{{\varphi _l}}

\newcommand{\elll}{{z_{s,l}}}

\newcommand{\ica}{{\mathcal I}}

\begin{document}

\title[Higher order statistics in the annulus square billiard]
{Higher order statistics in the annulus square billiard: transport and polyspectra}

\author{L Rebuzzini $^{1,2}$, R Artuso $^{1,3}$}

\address {$^1$ Center for Nonlinear and Complex Systems and
Dipartimento di Fisica e Matematica, Universit\`a dell'Insubria, 
Via Valleggio 11, 22100 Como, Italy.}
\address {$^2$ Istituto Nazionale di Fisica Nucleare, Sezione di Pavia, 
Via Ugo Bassi 6, 27100 Pavia, Italy.}
\address {$^3$ Istituto Nazionale di Fisica Nucleare, Sezione di Milano,
Via Celoria 16, 20133 Milano, Italy.}

\ead{laura.rebuzzini@uninsubria.it}

\begin{abstract}
Classical transport in a doubly connected polygonal billiard, i.e. the 
annulus square billiard, is considered.  
Dynamical properties of the billiard flow with a fixed initial direction are analyzed by 
means of the moments of arbitrary order of the number 
of revolutions around the inner square, accumulated by the particles during the evolution.  
An ``anomalous" diffusion 
is found: the  moment of order $q$ exhibits an algebraic growth in time 
with an exponent different from $q/2$, like in the normal case.  
Transport features are related to spectral properties of the system, which are reconstructed by Fourier 
transforming time correlation functions.    
 An analytic estimate for the growth exponent  of integer order moments is derived as a 
function of the scaling index at zero frequency of the spectral measure, associated 
to the angle spanned by the particles.  The $n$-th 
order moment is expressed in terms of a multiple-time correlation function, depending
 on $n-1$ time intervals, which is shown to be linked to higher order 
 density spectra (polyspectra),  by a generalization 
 of the Wiener-Khinchin Theorem. Analytic results are confirmed by numerical simulations.
\end{abstract}


\maketitle

\section{Introduction.}
\label{intro}
A  fundamental problem in statistical mechanics is to 
understand how the reversible microscopic dynamics of particles may produce macroscopic 
 transport phenomena, which are described by irreversible laws such as 
 diffusion equation or 
 Fourier law.  
In particular, the last few years have witnessed a large debate whether chaos at a microscopic level is a 
 necessary ingredient to generate realistic macroscopic behaviour \cite{decohvv,vu}. 
 In this paper we will be concerned with
 transport (diffusion) properties; for this purpose,  billiards 
  represent a class of dynamical systems ideally suited, as they allow both
 theoretical considerations and extensive numerical simulations: 
 transport is typically studied by lifting the billiard table on
 the plane, like in the case of the periodic Lorentz gas  \cite{decohvv,aacg,lerobe,ekmejia,mejia}. 
  We also remark their physical significance 
as simplified models to study energy or mass transport in realistic systems, such as 
fluids,  nanodevices,  electromagnetic cavities,  
optical fibers and low density particles in porous media \cite{jepbathia,davis,heric,klages}. 

While the origin of normal diffusion (i.e. when the mean square displacement of the particles 
grows asymptotically linearly in time) is well understood in fully chaotic billiards \cite{aacg,bor}, 
in recent years, great interest has focused 
on the study of transport in polygonal billiards, which are characterized by the 
absence of hyperbolicity  and 
dynamical chaos, in the sense of exponential divergence of nearby initial trajectories. 
Indeed in polygonal billiards, 
all the Lyapunov exponents and Kolmogorov-Sinai entropy vanish and  the 
dispersion of initially nearby orbits is polynomial.  
Nevertheless polygonal billiards may give rise to a wide 
range of transport regimes, extending from normal to ``anomalous" diffusion, in which 
the r.m.s. displays a non linear dependence on time  \cite{alonso1,lcw01,alonso2,rond,sand}. 
Some necessary conditions for the occurrence of normal transport in periodical 
billiard chains have been singled out: vertex angles irrationally related to $\pi$, 
absence of parallel scatterers, existence of an upper bound for the free path length between 
 collisions \cite{sandth}.

Particle 
dynamics in billiards can be described in terms of an invertible flow in 
continuous time or, equivalently, of an invertible map connecting two 
collision points, which corresponds to  
a unitary evolution ruled by the Koopman operator. 
The
main features of the dynamical transport are related  to 
 ergodic properties of the system, which can be formulated in terms of 
 properties of the 
spectrum of the Koopman operator.   
 Ergodic properties of polygonal billiards have been extensively tested by 
 the analysis of the decay of correlations, such as  
  mixing in the triangle with irrational angles \cite{caproz} and weak-mixing as the 
  maximal ergodic property in the right-triangle with irrational acute angles \cite{artcg}.   
The billiard flow in a typical polygon is ergodic, nevertheless the subset of 
{\it rational} polygonal billiards, in which 
all the angles between the sides are rational multiples of $\pi$, possess weaker ergodic properties. 
    Some exact results are known for this subset: 
  the billiard flow is not ergodic, 
because of a finite number of possible directions for a given initial condition, and it can be 
decomposed into directional flows, which are ergodic but not mixing, for a generic choice of 
  the direction and for almost all initial conditions  \cite{gut1,gut2}. 

\begin{figure}
\begin{center}
 \includegraphics[width=8cm,angle=0]{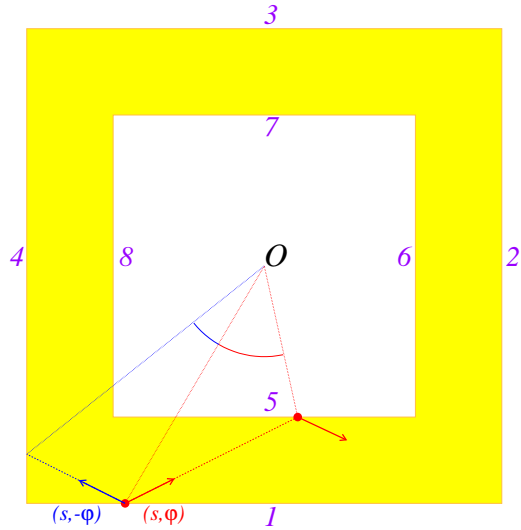}
 \caption{(Colour online) Billiard table for $l_1/l_2=\pi /2$. 
  A portion of a trajectory of a particle with initial condition $(s,\varphi)$,  
  evolving under $B_\varphi^t$ 
  with $t\geq 0$, is shown; $(s,-\varphi)$ is the "time reversed" initial condition.}
\label{bil}
 \end{center}
\end{figure}

The model considered in this paper is a 
 doubly connected rational billiards, called the annulus square billiard, which 
  was also studied in a quantum dynamical context \cite{berry,liboff}. 
  The billiard table, shown in figure \ref{bil}, is formed by the plane region included in between 
 two concentric squares with sides of different length.  Arithmetical properties of the 
 ratios between the sides of the two squares and between the components of the 
 velocity vector of the particle determine different dynamical and spectral features, which allow a 
 classification of the system into different subclasses \cite{agr}, 
 reviewed in section \ref{pro-ms}.   
 We inspect the angle (and its absolute value), accumulated by a particle 
 revolving around the hole, and provide a complete characterization of the transport 
  process by 
 the analysis of moments of arbitrary order
 of  this observable. The statistical approach is supported by a spectral analysis 
 of the system: the interdependence between dynamical and spectral properties leads to 
 a theoretical prediction for the exponents of moments of integer order.  
 We restrict to the class of billiards in which the ratios of the sides and velocity components
  are both irrational;  in this case, 
 for typical values of the parameters, the billiard flow, in a fixed direction, is weakly-mixing and 
 not mixing, entailing the presence of a singular continuous component in the spectrum, 
 namely supported on a set of zero Lebesgue measure \cite{agr}.  
 By assuming that the singularities of the spectral measure are of H\"older type \cite{hof,grac},  
 we derive an analytical relation connecting the exponents of the 
 algebraic growth in time of integer moments with the scaling index at zero frequency 
 of the spectral measure, associated to the angle. The formula is 
 exact for even-order moments, while it is an upper bound for 
 odd-order moments. 
 This result generalizes the one obtained in \cite{agr} for the mean square displacement
  (i.e. second order moment). Furthermore, analysis of higher order statistics 
  demands the introduction, in the context of dynamical system, of basic concepts  
  as multiple-time correlation functions and polyspectra, which are rather 
 familiar in signal analysis \cite{sim1}. 
  The theoretical relation for integer order moments 
 is tested by numerical simulations, which 
  moreover suggest that a similar estimate  
 holds for moments of arbitrary order: 
 the $q$-th order moment grows in time with an exponent $\nu q$; 
 $\nu$ is a constant function of the scaling index of the spectral measure in 0, which ranges 
 between $\nu =1/2$ (``normal" diffusion) and $\nu=1$ (``ballistic" transport). 

A more detailed characterization of the anomalous diffusion process has 
recently been considered in many papers about diffusion in periodical 
 polygonal billiard channel, in which the polygonal scatterers form a 1-dimensional 
 periodical array \cite{alonso1,lcw01,alonso2,sand,armstead}. These studies constitute 
 a further motivation to the analysis of transport in the square annulus billiard; 
 indeed, to examine the dynamics of particles winding around the inner square is 
  equivalent to consider the transport in a ``generalized Lorentz gas", i.e. 
  an extended system, obtained by periodically repeating the elementary cell in both directions. 
 In periodic channels,  two different 
 behaviours are identified:  ``weak" anomalous diffusion, when the exponent of the algebraic  
 growth of  moment of order $q$  is a linear function of the order, namely 
 $\nu q$ with $\nu =const. \not = 1/2$  for $\forall q$, and ``strong" anomalous diffusion, 
 when the exponent is a non trivial function of the order, i.e. 
 $\nu (q) q$ \cite{vulp, fmy01}.  
  In some polygonal chains, 
  a transition from normal to anomalous 
 diffusion has been found  in cases with  
 an infinite horizon, i.e. when the particle may travel without colliding 
 with the walls, or, when the particle can  propagate arbitrarily far by reflecting off only on parallels 
 scatterers \cite{rond,sand}. A value $q_{cr}$ exists that discriminates between the two 
 regimes with different slopes of the exponent versus $q$.   
 This phase transition physically corresponds to a change in the balance between 
 the ballistic and diffusive trajectories in the ensemble averages and infers the absence of a 
 unique scaling law for the probability density of particles, which does not 
 relax at long times into a self-similar profile; this kind of processes are therefore 
 classified as  `` weak" self-similar processes.  In some other cases, 
 such as in the ``zigzag model" \cite{sandth,vulp}, the phase transition is not present;  
 this seems to be related to the geometry of the polygonal channel and, in particular, 
 to the number-theoretical properties of the angles.

The paper in organized as follows. The billiard model is described in section 
\ref{model} and its basic spectral properties are reviewed. In section \ref{link} we 
 introduce the phase observable which identifies the diffusive process and 
 we developed the spectral analysis of the signal produced by time 
 evolution. 
  We then consider arbitrary order moments of the observable;  
the crucial issue is to get informations on
 the scaling function, which determines the algebraic growth of the moments 
   for long times. 
  From the analytic point of view we restrict to  integer order moments: higher order moments are expressed in terms of 
multiple-time correlation functions and polyspectra, which are introduced in section \ref{hstat}. 
The analytic relation between spectral and 
  dynamical indexes  is derived  in section 
  \ref{result} and numerically confirmed in section 
  \ref{num}.   

\section{The model.} 
\label{model}

\subsection{Description of the billiard table}

We consider the dynamics of a classical point particle in 
a doubly connected rational polygonal billiard.  
The billiard table, shown in figure  \ref{bil}, is delimited by two concentric squares 
with parallel sides of length $l_1$ and $l_2$ ($l_1>l_2$). 
The particle moves freely inside the billiard, colliding elastically with the boundary; the (preserved) modulus of the velocity is taken equal to $1$.  The accessible phase space is 3-dimensional and the 
representative point is $(x,y,\theta)$: $(x,y)$ are the Cartesian coordinates 
of the particles and 
$\theta$ is the angle of the velocity, measured counterclockwise from 
the positive $x$-axis. The billiard flow, denoted 
by $T^t$,  preserves the Lebesgue measure 
$\dd x\dd y\dd\theta /(2\pi (l_1^2-l_2^2))$. 

We may also consider the discrete time dynamics, induced by successive collisions. 
Each collision point can be  parametrized by 
a coordinate $s$ along the boundary 
and an angle $\varphi$, between the 
outgoing direction and the inner normal to 
the boundary; we take $\varphi$  
positive if the outgoing velocity is given by a counterclockwise rotation to the inner normal.  
By denoting $L=2(l_1+l_2)$, the 
arclength $s\in [-L,L)$  takes 
the value  
$s=-L$ in the left corner of side 1 and  increases by  
moving counterclockwise  along the sides of the billiard, labeled by 
increasing numbers (see figure  \ref{bil}). The
corresponding phase space is called  ${\mathcal M}=\{  (s,\varphi); s\in [-L,L) ,  \varphi\in [-\pi/2,\pi/2) \}$ and the Birkhoff-Poincar\'e map is denoted by
$B^t$ ($t\in {\ZM}$); 
  the map $B^t$ preserves the measure 
     $\dd \Omega(z) =\cos \varphi \dd \varphi \; \dd s\; / (4L)$ on ${\mathcal M}$.
     
     Dynamics on such billiards is strongly influenced by number--theoretical properties of $l_2/l_1$ and of $\tan \varphi$; in particular 
     we may distinguish three classes \cite{agr}: $(i)$ {\it class} 1:  $l_2/l_1, \tan \varphi\in \QM$; 
$(ii)$ {\it class} 2:  $l_2/l_1\in \QM$, $\tan \varphi\in {\RM -\QM}$; 
$(iii)$ {\it class} 3:  $l_2/l_1\in \RM -\QM$, $\tan \varphi\in {\RM -\QM}$. 
In this paper we restrict to {\it class} 3.

\subsection{Review on basic spectral properties}
\label{pro-ms}

As all the internal angles are rational multiples of $\pi$, it is fairly easy to realize that the flow on the global phase space, or the mapping $B^t$ on ${\mathcal M}$, can never be ergodic.  However   $T^t$ 
can be  decomposed into the one-parameter family of directional flows at fixed $\theta$, 
$T^t_{\theta}$, whose dynamics is not trivial:
in particular,  $T^t_{\theta}$ is ergodic for almost all $ \theta\in [0,  \pi /2)$ and never
 mixing  \cite{gut1,gut2}. It is believed that the
weak-mixing property should be enjoyed by a ``generic" subset of directional flows, and numerical evidences for {\it class 3} are provided
in \cite{agr}.

Analogously to the billiard flow, the map $B^t$ can be decomposed into a
one-parameter family of components 
 $B^t_{\varphi}$ along the 
fixed direction $\varphi$ (with 
$0 \leq \varphi \leq \pi /2$).  
Once the initial outgoing angle is set, only the angles $\pm \varphi, \pm (\frac \pi 2 -\varphi)$ can be met along the trajectory, so the phase space for any fixed foliation, i.e. ${\mathcal M}_\varphi$, is a fourfold replica of $[-L,L)$: its points are denoted by $z=(s,\pm \varphi _j)$ with 
 $\varphi _j=\varphi -j\pi /2$ and $j=0,1$. The phase average of an observable $f: {\mathcal M}_\vz \to \CM$ can be 
 written explicitly as:
\begin{equation}
\label{ph-ave}
\int _{{\mathcal M}_\vz} \dd \Omega (z) \; f(z)  = 
\frac {1}{8 L} \int _{-L}^{L} \dd s \; 
\sum _{j=0}^1 \cos \vj \left\{ f(s,{\varphi _j})+ 
 f(s,-{\varphi _j})\right\}.
\end{equation}

Spectral properties of the system can be formulated in terms of the  asymptotic behaviour of 
the correlation functions. 

Ergodicity of $B^t_\varphi$ implies that the 
time-averaged correlation function of a generic (real) phase function $f : {\mathcal M}_\vz\to \RM$: 
\begin{equation}
\label{ctime}
C^{time}_f (t,z) =\lim _{T\to +\infty} \frac 1{2T} 
\sum _{t' =-T}^{T-1} f (B^{t' +t}_\varphi z) f (B^{t'}_\varphi z), 
\end{equation}
and the 
phase-averaged correlation function:
\begin{equation}
\label{cphase}
C^{ph}_f (t) =\langle f | U^t f\rangle=\int _{{\mathcal M}_\vz} \dd \Omega (z) \; f (B_\vz^t z) f(z), 
\end{equation}
 coincide for $\Omega$-almost every $z\in {\mathcal M}_\vz$.  
 The unitary operator $U$ in (\ref{cphase}) is the Koopman operator, associated to the 
 map $B^t_\varphi$.

\begin{figure}
 \begin{center}
    \includegraphics[width=8cm,angle=0]{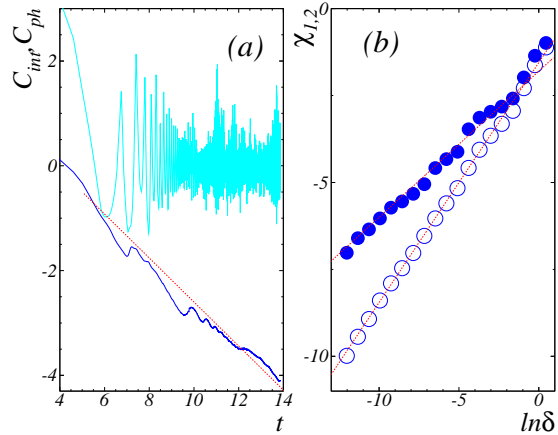}
  \caption{(Colour online) (a) Phase averaged (upper line) and integrated (lower) correlation functions 
of the angle $\xi$, spanned by the radius vector, as a function of time $t$  
 (measured in numbers of collisions).
The billiard table belongs to {\it class 3} with parameters:  
 $l_1/l_2=\pi/2$ and  $\tan\varphi=\pi/4$. 
  The dotted line, which fits $C^{int}(t)$, has a slope equal to $-D_2$.
     (b) Estimates of  generalized dimensions 
$D_1$ (empty symbols) and $D_2$ (full symbols) of the spectral measure, 
associated to $\xi$, 
for the same parameters. The dimension estimates are 
given by the slopes of straight lines: $D_1=0.69\pm 0.01$ and $D_2=0.42\pm 0.02$. 
Details are explained in section \ref{num}.}
\label{co-d1d2}
 \end{center}
\end{figure}

As directional flows or mappings are not mixing, correlation functions 
(of zero mean observables) do not decay to zero 
in the asymptotic limit; while the (conjectured) weak-mixing property guarantees that integrated correlations:
\begin{equation}
\label{cinte}
C^{int}_{f} (t) =\frac 1t \sum _{s=0}^{t-1} \left| C^{ph}_f (s)\right|^2
\end{equation}
vanish as $t \to \infty$.
Phase-averaged correlation function and integrated correlation function 
 of the angle $\xi$, spanned by the radius vector of a particle 
 between two successive collision points, 
  are shown in figure \ref{co-d1d2}(a) 
 for a generic billiard belonging to {\it class 3}: while $C^{ph} (t)\not \to 0$ (upper line), 
 $C^{int}_{f} (t)$ 
 decays to zero polynomially (lower line). 

By using the spectral decomposition of the Koopman operator $U$, we may rewrite (\ref{cphase}) as 
\begin{equation}
\label{cphase-ms}
C^{ph}_f (t) = \frac {1}{2\pi} \int _{-\pi}^{\pi}\dd \mu _f (\theta) e^{i\theta t},
\end{equation}
 which provides a direct link between the autocorrelation function of an observable $f$ 
 and the associated 
 spectral measure. If (in the complement of constant functions) the spectral measure is absolutely continuous, the 
 system is mixing.  On the other side, weak mixing is equivalent to an empty point spectrum, 
 apart from the eigenvalue 1.   Therefore, 
 the weak mixing property, without the stronger mixing property, entails the presence 
of a singular continuous component of the spectrum of the Koopman operator. 

Owing to the presence of a non empty point spectra, 
weakly mixing property is ruled out in the almost integrable cases, namely for $l_1/l_2 \in {\QM}$ 
({\it class 1} and {\it class 2});  we will not consider such cases in the present work.

In \cite{agr}, the occurrence of a singular continuous component of the 
 spectrum   in billiards of {\it class 3} was inferred by looking at 
 scaling properties of the spectral measure,
obtained 
 by finer and finer numerical inversion of (\ref{cphase-ms}). 
 In particular a multifractal analysis yields a nontrivial spectrum of generalized dimensions, with a Hausdorff dimension $D_1$ less than 1
 and a correlation dimension $D_2$ which rules the power-law decay of integrated correlations 
 \cite{gei,hol}. 
 An example is shown in figure \ref{co-d1d2}(b), for the spectral measure associated to the angle $\xi$. 
  As we will show in following sections, the local scaling properties of the spectral measure are 
 even more relevant  in connection with transport properties. For billiards in {\it class 3},  
  a nontrivial scaling of the spectral peaks near the zero frequency is found (see figure \ref{ms-alp}(b)), 
  in opposition to the almost integrable case, in which a non empty pure point spectrum 
  is marked by not scaling deltalike peaks at different resolutions \cite{agr}.

\section{Higher order statistics}
\label{link}

\subsection{Dynamical quantities and spectral analysis}

The dynamics inside the billiard table in figure \ref{bil} is equivalent to the dynamics 
of the particles in a two-dimensional infinite periodic lattice with square obstacles, i.e. 
in a generalized Lorentz gas, recently examined in \cite{alonso1,lcw01,alonso2, sand,armstead}. 
The unfolded system is obtained by reflecting the elementary cell and the segment of a trajectory 
at each collision point with the external square. 
  Instead of considering particle diffusion along the channels of the extended system, we examine
 the transport generated by billiard trajectories revolving around the 
inner square obstacle. 

 For this purpose, the natural observable is the 
angle $\xi (z)$, spanned by the radius vector, joining the center of the 
billiard $O$ with the collision point $z$, when the particle is moving 
 from $z$ to $B_\vz z$; it is assumed positive when counterclockwise.
 The total angle accumulated by a single particle up to time $t$ is: 
\begin{equation}
\label{teta}
\Xi (z,t)=\sum _{s=0}^{t-1} \xi (B_\vz^s z); 
\end{equation}
 $\Xi (z,t)/ (2\pi)$ gives the number of revolutions 
 completed by a trajectory up to the time $t$. 
 
 In \cite{agr}  the 2-nd order moment $\sigma ^2(t)$ of 
$\Xi (z,t)$, 
namely the r.m.s. number of revolutions, was examined; for generical values of the parameters 
in {\it class} 3, 
an ``anomalous diffusion" was found, marked by an 
algebraic growth of $\sigma ^2$ in time with an exponent ranging between 1 (normal diffusion) 
and 2 (ballistic transport); this exponent has shown to be related 
to the zero-frequency 
scaling index of the density power spectrum associated to 
the observable $\xi (z)$.  In this paper we extend the analysis to moments 
of arbitrary order. 
  

Since $\xi (B_\varphi^t z)$ is a ``power signal", i.e. $\{ \xi (B_\vz ^t z) \}_{t\in \ZM }
\not\in \ell^2 (\ZM)$, as 
a function of time $t$,  but  $W=\lim _{T\to\infty }\frac {1}{2T}\sum_{t=-T}^{T-1}\xi^2 
 (B_\varphi^t z)\equiv C^{time}_\xi (0 ; z) <+\infty$, we analyse the trajectory 
of a particle on a finite time interval: $-T\leq t\leq T$ (with $T$ positive integer, 
  i.e. $T\in \NMp$).
  Spectral analysis will involve Fourier transform of the signal  $\xi (B_\vz ^t z)$ on finite portions of trajectories.  
   
%
Hence we call $\hat \xi_T$ the
partial sum of the Fourier series of $\xi (B_\vz^t z)$:
\begin{equation}
\label{fou}
\fl \hat \xi_T (\theta ;z)=\left( {k}_T \ast \hat \xi \right)(\theta ;z) 
=\sum _{t=-T}^{T-1} \xi (B_\vz^t z) e^{-it\theta}=\rho _T (\theta ;z)e^{i\phi _T (\theta;z)}, 
\end{equation}
where $\ast$ denotes convolution and ${k}_T (x)=\sum _{t=-T}^{T-1}e^{-ixt}=e^{i\frac x2} \left( \frac {\sin Tx}{\sin \frac x2}\right)$.
The upper limit of the sum in (\ref{fou})
 is determined by the fact that $\xi (B_\vz^T z)$ is not defined.

\begin{figure}
 \begin{center}
  \includegraphics[width=8cm,angle=0]{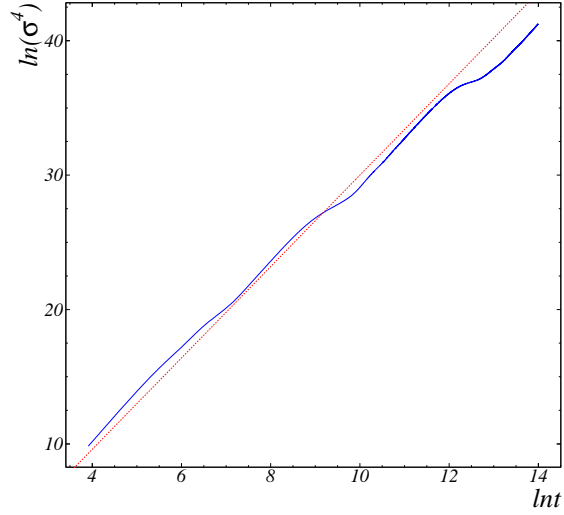}
  \caption{(Colour online) Fourth order moment of $\Xi (z,t)$ vs time in logarithmic scales, 
  for the same data of figure \ref{ms-alp}.  
  The slope of the straight line (= 3.4) is given by the theoretical prediction (\ref{result-n}) with 
  $n=4$ and $\ba (0)=0.3$.}
\label{mo4}
 \end{center}
\end{figure}

 $\left\{ \hat \xi_T (\theta ;z)\right\}_{T\in \NMp}$ is a sequence of continuous, bounded 
  and periodic complex function of the frequency 
  $\theta \in \TM$;  $\TM = \RM /(2\pi \ZM)$ denotes the 1-dimensional torus.   
As $\xi (B_\vz^t z)\not\to 0$ for $t\to \infty$, 
the limit $T\to \infty$ of 
the sequence of partial sums  may not be a function in ordinary sense;  as 
$|\xi (t;z)|\leq A$ $\forall t\in \ZM$ and $0<A<\pi$, from the theory of Fouries series, it follows  
that  $\left\{ \hat \xi_T (\theta ;z)\right\}_{T\in \NMp}$ converges weakly for $T\to \infty$ to 
a tempered distribution of period $\TM$. 

Since $\xi$ is a real-valued function, 
$\hat \xi_T^* (\theta ;z) =\hat \xi_T (-\theta ;z) $ and the absolute value and the phase of $\hat \xi_T$
 are respectively even and odd functions of $\theta$  (mod $2\pi$) :
 \begin{eqnarray}
  \label{even}
& & \rho _T (-\theta;z)=\rho _T (\theta;z) =\rho _T (|\theta |;z)  \\
\label{odd}
& & \phi _T (-\theta;z)=-\phi _T (\theta;z).
\end{eqnarray}
In particular 
$\hat \xi_T(0;z)=\rho _T (0;z)$,  $\phi _T (0;z)=0$; while,  from periodicity 
and from (\ref{odd}), it follows that $\phi _T (-\pi;z)= \phi _T (\pi;z)=0$.

%
%
%
By looking at forward trajectories starting at $(s, \varphi)$ and their time reversal, generated by 
the ``backward" (inverse) operator $B_\vz^{-1}$ 
on $(s, -\varphi)$, we can verify that the following identity holds:
%
\begin{equation}
\label{timeinv}
\xi (B_\vz^t (s,-\varphi))= - \xi (B_\vz^{-t-1}(s,\varphi)).
\end{equation}
This identity 
induces the following property of the partial sums:
\begin{equation}
\label{fou-ti}
\hat \xi _T (\theta; (s,-\varphi))= - e^{i\theta }\hat \xi _T^* (\theta; (s,\varphi)),
\end{equation}
which is equivalent to
$\rho_T (\theta; (s,-\varphi))=
\rho_T (\theta; (s,\varphi))$ and $\phi_T (\theta; (s,-\varphi))=\theta+\pi -\phi_T (\theta; (s,\varphi))$ 
(mod ($2\pi$)).


\subsection{Higher order moments and correlations} 

 To get a more detailed picture of anomalous transport, 
 we introduce the higher order statistics of $\Xi (z,t)$ and 
higher order spectral analysis of the phase variable $\xi (z)$. 
 
 While, from the numerical point of view, 
we may take arbitrarily real powers of the diffusing variable $| \Xi(z,t)|$,  
 in the theoretical analysis of sections \ref{link}, \ref{hstat} and \ref{result}, we restrict to 
integer order moments of $\Xi (z,t)$. Integer order  
 moments can be expressed in terms of multiple-time correlation functions, 
 related to higher order spectra.  

 The $n$-th order moment is given by:
\begin{equation}
\label{n-mo}
\sigma ^{(n)} (t) =\int _{{\mathcal M}_\varphi} {\dd}\Omega (z) \Xi ^n (z,t)
\qquad\quad n\geq 2.
\end{equation} 
The occurrence of anomalous diffusion, previously found for $\sigma ^2(t)$ \cite{agr}, 
is confirmed for moments of arbitrary orders; in figure \ref{mo4} the 
fourth-order moment is shown, 
as an example. The crucial issue is to get an estimate for the exponents of the 
algebraic growth in time of integer moments. 

By a generalization of (\ref{cphase}), we introduce a multiple-time 
phase-averaged correlation function of the observable $\xi$, 
which depends on $(n-1)$ time intervals:
\begin{equation}
\label{cphasen}
\fl C^{ph}_\xi (\vec t)=
\int _{{\mathcal M}_\varphi } \dd \Omega (z) \; \xi(z) \left( \prod _{l=0}^{n-1} \xi(B_\vz ^{t_l}z) \right), 
 \qquad \vec t =(t_1,t_2,\cdots,t_{n-1})\in \ZM^{n-1}.
\end{equation}

The $n$-th moment $\sigma ^{(n)} (t)$ can be expressed as a function of $C^{ph}_\xi (\vec t)$:
\begin{eqnarray}
 \label{n-mo-1}
\fl \sigma ^{(n)} (t) 
&=&  \sum _{t_0 =0}^{t-1}\sum _{t_1 =0}^{t-1}\cdots \sum _{t_{n-1} =0}^{t-1} 
 \int _{{\mathcal M}_\varphi}  \dd \Omega (z) \; \xi(B_\vz ^{t_0}z) \xi(B_\vz ^{t_1}z)  \xi(B_\vz^{t_2}z)   
 \cdots \xi(B_\vz^{t_{n-1}}z)\nonumber \\
\fl &=& \sum _{t_0 =0}^{t-1}\sum _{t_1 =0}^{t-1} \cdots \sum _{t_{n-1} =0}^{t-1}  
 C^{ph}_\xi (\vec t -t_0\vec I),
 \end{eqnarray}
 where $\vec I$ is the vector with all entries equal to 1;
note that the $n$-th order moment is related to a $(n-1)$-point correlation function.

\section{Higher order spectra analysis.}
\label{hstat}

\subsection{Multi-dimensional Wiener-Khinchin theorem}

The
 $(n-1)$-point correlation can be expressed in terms of a $(n-1)$-dimensional inverse Fourier 
transform of the $(n-1)$-th order spectral density distribution function
 (polyspectrum) on $\TM^{n-1}$ \cite{hosa,poly}:
\begin{equation}
\label{wk-n}
C^{ph}_\xi (\vec t\; )=\frac {1}{(2\pi)^{n-1}} \int_{\TMN}\ddb \vt \; 
  m_\xi (\vt  ) 
 e^{i \vt  \cdot \vec t}; 
\end{equation}
$ \TM^{n-1}$ is the $(n-1)$-dimensional torus, i.e. 
the Cartesian product of $(n-1)$ tori $\TM\times\TM\cdots\times\TM$ and 
$\vt  = (\theta_1,\theta_2,\cdots,\theta_{n-1})\in  \TM^{n-1}$. 
Equation (\ref{wk-n}) is a formal generalization of the Wiener-Khinchin Theorem
 to the multi-dimensional case. 
However, since for $n>2$, $ C^{ph}_\xi (\vec t\; )$ is not a positive definite 
function, according to Bochner's Theorem, $m_\xi (\vt  )$ is not  
a positive distribution and $m_\xi (\vt  )\ddb \vt$
cannot be interpreted as a probability measure on $\TM^{n-1}$ \cite{striz,abra}.  

An alternative version of the theorem involves the time-averaged correlation function:
\begin{eqnarray}
\label{wk-n-time-1}
C^{time}_\xi (\vec t ; z)&&\equiv
\lim _{T\to \infty}\frac 1{2T}\sum _{t=-T}^{T-1}
\xi(B_\vz^t z) \left (\prod _{l=1}^n 
\xi(B_\vz^{t+t_l}z)  \right) \\
\label{wk-n-time-2}
&&= \frac {1}{(2\pi)^n} \int_{\TMN}\ddb \vt \; 
  s_\xi (\vt ;z ) 
 e^{i \vt  \cdot \vec t}, 
\end{eqnarray}
with
\begin{equation}
\label{sp-avera}
m_\xi  (\vt  )=\int_{{\mathcal M}_\varphi} \dd
 \Omega (z) s_\xi (\vt  ;z). 
\end{equation} 
If the system is ergodic, $s_\xi (\vt ;z)=m_\xi (\vt )$, for $\Omega$-almost every $z\in 
{\mathcal M}_\varphi$.

\subsection{Polyspectra}

  We may obtain  $m_\xi(\vt)$  in terms of partial sums of  the 
Fourier series of $\xi (B_\varphi^t z)$. 

Firstly, we derive $s_\xi (\vt  ;z)$ from (\ref{wk-n-time-2}). 
Since we consider the trajectory 
of a particle on a finite time interval, 
  each signal $\xi(B_\vz^{t+t_l}z)$ in (\ref{wk-n-time-1}) is substituted by 
 $\chi _{\ica _T} (t+t_l)\cdot \xi(B_\vz^{t+t_l}z)$, where $\chi_{\ica _T}$ is the 
 characteristic function of the interval of integers ${\ica _T} \equiv \{ t\in \ZM ; -T\leq t\leq T-1\}$. 
 
 The square window of the signal is expressed by the  Fourier transform
\begin{equation}
\label{inv-tf}
{\chi}_{\ica _T } (t) \xi (B^t z)=\frac 1{2\pi}\int _{-\pi}^\pi \dd \theta \; \hat \xi_T (\theta;z) e^{i\theta t}, 
\end{equation}
which is obtained by the Convolution Theorem from (\ref{fou}).

The polyspectra $s_\xi (\vt ; z)$  (and thus $m_\xi (\vt )$) is
given as a limit of a sequence of 
 functions $\{ s_{T,\xi} (\vt ;z) \}_{T\in \NMp}$, which converges to $s_\xi (\vt ;z)$ in a 
 weak sense:  
\begin{equation}
s_\xi (\vt ; z) =  \lim _{T\to \infty}  s_{T,\xi} (\vt ; z)
\quad {\rm weakly}. \nonumber 
\end{equation}

By inverting (\ref{wk-n-time-2}) and substituting (\ref{inv-tf}), we get:
\begin{eqnarray}
\fl && s_{T,\xi} (\vt ; z) = \frac 1{2T}
  \sum _{t=-T}^{T-1}
\sum _{t_1=-T-t}^{T-t-1}
\cdots \sum _{t_{n-1}=-T-t}^{T-t-1}   e^{-i\vt\cdot \vec t}
\chi_{\ica _T } (t) \xi (B^t z)
\prod  _{l=1}^{n-1} \chi_{\ica _T } (t+t_l) \xi (B^{t+t_l} z)  \nonumber \\
\fl && \quad\qquad\;\;= \frac 1{2T(2\pi )^{n}} 
 \int _{\TM}\dd \theta \  \int _{\TM ^{n-1}} 
\ddb \vt '  \hat\xi_T(\theta;z) k^*_T(\theta +\Theta (\vt )) 
 \prod  _{l=1}^{n-1}  \hat\xi_T(\theta'_l ;z) k_T (\theta_l - \theta'_l ), \nonumber
\end{eqnarray}
where $\Theta (\vt )=\sum_{l=1}^{n-1} \theta_l $. 
By using that $k_T(x)$ (and $k^*_T(x)$) is an approximate identity, i.e. 
$\lim _{T\to\infty}\int _{\TM} \dd \theta k_T (\theta -\theta ')f(\theta)=2\pi f (\theta ')$ 
for all continuous functions $f$ on $\TM$, the final expression of $s_\xi (\vt ;z)$ is obtained: 
\begin{equation}
\label{poly-n-ap}
s_\xi (\vt  ;z)= \lim _{T\to \infty}\frac {1}{2T}
\left( \prod_{j=1}^{n-1} \hat\xi_T(\theta_j;z)\right)
\hat\xi^*_T\left(\Theta (\vt ) ;z\right ).
\end{equation}
$m_\xi (\vt )$ is derived by applying phase averages (\ref{sp-avera}) to both sides of (\ref{poly-n-ap}).
 The result is consistent with analogous formulas in \cite{hosa,pdom3}. 

We introduce the notation: $\Gamma _T (\vt ; \elll )= \prod _{j=1}^{n-1} \rho_T(\theta_j;\elll )$ 
and $\Phi_T(\vt ;\elll )=\sum _{j=1}^{n-1}  \phi_T(\theta_j ; \elll )$, 
with $\elll =(s,\varphi_l)$ and $\varphi_l =\varphi -l\pi /2$ ($l=0,1$).

By evaluating the phase average with (\ref{ph-ave}) and by making use of the property (\ref{fou-ti}), 
 we get:
\begin{eqnarray}
\label{mtn}
&& m_\xi  (\vt  )= \lim _{T\to \infty} m_{T,\xi}  (\vt  ) \quad {\rm weakly} \nonumber \\
&& m_{T,\xi}  (\vt  ) = \frac 1{4L} \int _{-L}^L \dd s\sum _{l=0}^1 \cos \vl \; 
  \cdot M_{T,\xi} (\vt ;\elll); 
\end{eqnarray}
 for $n$ odd integer:
\begin{equation}
\label{m-odd}
\fl M_{T,\xi}(\vt  ;\elll )=\frac {i}{2T}
\Gamma _T (\vt ; \elll )\rho_T (\Theta (\vt  );\elll )\sin \left( \Phi _T (\vt ; \elll)- \phi_T (\Theta (\vt ) ;\elll ) 
 \right); 
\end{equation}
for $n$ even integer: 
\begin{equation}
\label{m-even}
\fl M_{T,\xi}  (\vt  ; \elll )=\frac {1}{2T}
\Gamma _T (\vt ; \elll )\rho_T (\Theta (\vt );\elll )\cos \left(  \Phi _T (\vt ;\elll )   - \phi_T (\Theta (\vt ) ;\elll ) 
\right).
\end{equation}
Explicit formulas for the first few orders are given in \ref{cor-pol}.

Higher order spectra $m_\xi  (\vt  )$ possess different symmetry properties. 

From (\ref{poly-n-ap}) and since $\hat \xi^*_T (\theta ; z)=\hat \xi_T (-\theta ; z)$,  a
conjugate symmetry property holds for $m_\xi (\vec \theta)$:
\begin{equation}
\label{mcc}
m_\xi ^*(\vt  )=m_\xi  (-\vt  );
\end{equation}
this condition guarantees the reality of multiple-time correlation $C_{\xi}^{ph}(\vec t)$, 
expressed by (\ref{wk-n}).

In particular, from (\ref{m-odd}), (\ref{m-even}),  (\ref{even}) and (\ref{odd}), 
 we get:
\begin{eqnarray}
\label{m-odd-1}
& m_\xi  (-\vt  )=-m_\xi (\vt  ) \qquad
\qquad & {\rm for}\; {\rm odd} \;  n\\
\label{m-even-1}
& m_\xi (-\vt  )=m_\xi (\vt   ) \qquad
\qquad & {\rm for}\; {\rm even}\;  n.
\end{eqnarray}

For odd $n$ (\ref{m-odd-1}) entails that the multiple integral of $m_\xi (\vt )$ on $\TM ^{n-1}$ is 
null, because the contributions of  hyperoctants associated to $\vt $
 and $-\vt$ cancel each other.  For even $n$ instead, owing to (\ref{m-even-1}), 
 these contributions are equal and the 
 integration domain can be reduced from $2^{n-1}$ hyperoctants to $2^{n-2}$ hyperoctants.
 
The multiple-time correlation function $C^{ph}_\xi(\vec t  )$ posseses further 
symmetry properties, which reflect to symmetries of 
$m_\xi  (\vt  )$ and allow to further reduce the integration domain in (\ref{wk-n}) 
to the ``so-called" principal domain \cite{sim1,pdom}. These symmetry conditions are 
reviewed in \ref{sim} for the bispectrum and trispectrum, i.e. $n=3$ and $n=4$.

\subsection{Single frequency case: small frequency asymptotics of the spectral measure} 

\begin{figure}
 \begin{center}
 \includegraphics[width=8cm,angle=0]{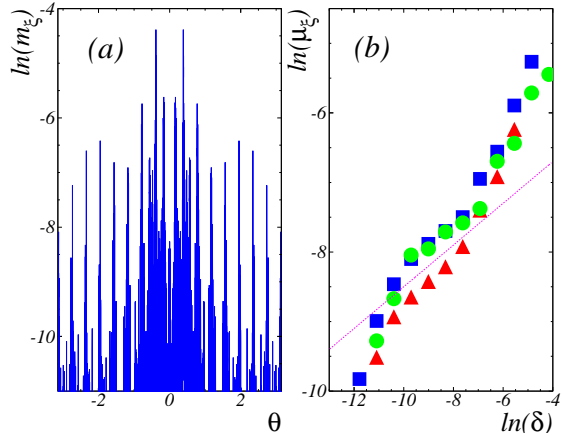}
 \caption{(Colour online) (a) Numerical reconstruction of the 
 density power spectrum $m_\xi (\theta )$ associated to the angle $\xi$ with a 
 resolution $\Delta \theta =2\pi / 2^{14}$,  for  $l_1/l_2=\pi/2$ and 
 $\tan\varphi=\pi/4$.
(b) Mass contained in small intervals of radius $\delta$ centered at $0$ vs $\delta$. 
    The slope of the straight line gives the  H\"older exponent at $\theta=0$, 
    i.e.  $\ba (0) =0.30$.
  Different symbols refer to different resolutions of $m_\xi(\theta)$, as explained in section \ref{num}. }
\label{ms-alp}
 \end{center}
\end{figure}

In section \ref{result} we make use of the asymptotic behaviour of polyspectra for 
vanishing frequencies.  Since (\ref{poly-n-ap}) and (\ref{mtn}) are expressed in terms 
of products of one-variable functions, in this section we focus on 
 the single frequency case, namely $n=2$. 

The 
correlation functions (\ref{wk-n}) and (\ref{wk-n-time-2}) are expressed by 1-dimensional 
integrals; $s_\xi (\theta ;z)$ is the density power spectrum of the signal, obtained by a weak limit 
of the sequence of  functions $\{ s_{T,\xi} (\theta ;z\}_{T\in \NMp}$:
 \begin{equation}
\label{dps}
 s_{T,\xi} (\theta ;z)= \frac {1}{2T} \hat \xi _T (\theta ; z)\;
\hat \xi ^*_T (\theta ; z)=\frac {1}{2T} 
\rho _T ^2 (\theta ;z).
\end{equation}

 By comparing (\ref{wk-n}), for $n=2$, with (\ref{cphase-ms}) we have
\begin{equation}
\label{ps-ms}
\dd \mu _\xi (\theta) =m_\xi (\theta )\dd \theta,
\end{equation}
which has to be interpreted in a strict distributional sense, 
as in our case $m_\xi(\theta)$ will be a singular object. In billiards of {\it class 3}, 
as explained in section \ref{pro-ms}, 
the absence of mixing excludes occurrence of purely absolute continuous spectrum and  implies that  
 the correlations do not decay to zero as $t\to\infty$; therefore, owing to 
  Riemann-Lebesgue Lemma, 
$m_{\xi} (\theta)$ is not an integrable function of $\theta\in {\mathbb T}$. 

In \cite{agr} it was pointed out that the singularity of the measure at $\theta=0$ is essential in order to get an anomalous second moment of the transporting variable; in particular the indices that quantify such singularities are the critical H\"older exponents $\ba (\theta)$, which are defined for each point $\theta$ in the support of the measure as  
\begin{equation}
\label{ass}
\limsup_{\delta \to 0} \mu_\xi({\mathcal I}_\delta (\theta))\cdot \delta^{-\alpha}=\left\{
\begin{array}{ll}
0 & \alpha < \ba (\theta) \\ \infty  & \alpha> \ba (\theta)
\end{array}\right. ,
\end{equation}
where ${\mathcal I}_\delta (\theta) =[\theta -\delta, \theta +\delta ]$ is an 
interval of width $2\delta$, centered in 
$\theta$. 
The measure is uniformly $\ba$-H\"older continuous (U$\ba$H) in an interval 
${\mathcal I}_\Delta (\tilde \theta)$,  centered in a particular point $\tilde{\theta}$, 
 if a positive constant $c$ exists  such that the mass 
 $\mu _\xi ({\mathcal I}_\delta (\theta) )\leq c\delta ^{\ba (\tilde \theta)}$ for every 
interval ${\mathcal I}_\delta (\theta)\subset {\mathcal I}_\Delta (\tilde \theta)$. 

 Smaller values of $\ba$ correspond 
 to stronger singularities of the spectral measure. The more interesting case 
 is when $0\leq \ba \leq 1$; in the following, we will consider $\ba (\tilde \theta)$ 
 varying within this range. The values $\ba =0$ and $\ba \geq 1$ correspond 
 to a discrete and absolutely continuous component of the spectral measure,  respectively; 
 if $\ba (\tilde \theta )=0$,  $\mu_\xi (\{ \tilde \theta \})>0 $ and 
 if $\ba (\tilde \theta ) \geq 1$, the measure is continuous and differentiable in 
${\mathcal I}_\Delta (\tilde \theta)$ and the derivative $m_\xi (\theta)$, being  
an integrable function on ${\mathcal I}_\Delta (\tilde \theta)$, is the density of the measure, 
with respect to Lebesgue measure. 
  According to numerical approximations of the spectral measure, billiards belonging to 
 {\it class} 3 are characterized by values of $\ba$  in the range $0<\ba <1$; in figure \ref{ms-alp} a 
 typical case is shown, in which $\ba (0)= 0.3$. 
 This is consistent with the presence of a singular 
 continuous component of the spectrum; however, an exponent $\ba \in (0,1)$ 
 is not a sufficient condition to have either a continuous \cite{agr} or singular 
 continuous part \cite{hof} of the measure.
 
 In section \ref{result} we will use a relationship which connects the
H\"older exponent of the measure at some point $\tilde \theta$ to the asymptotic 
behaviour as $T\to +\infty$
 of the sequence 
$\{ m_{T,\xi}(\theta ) \}_{T\in \NMp}$;  it is based on the equality $m_{T,\xi} (\theta ) = 
\frac 1{2\pi} \int _{\TM} \dd \mu _\xi (\theta ') K_{T} (\theta -\theta ')$  with 
$K_{T} (x)= (  {\sin Tx} / {\sin(x/2) } )^2 /2T$. 
This relationship 
is derived in  \cite{hof} and reviewed in \ref{kernel-hof}.   

If the measure is U$\ba$H in an interval ${\mathcal I} _\Delta(\tilde{\theta})$, 
 then there exist a positive constant $D$ and $\bar T =\bar T (\Delta , \theta )$ 
 such that for  $T>\bar T$: 
\begin{equation}
\label{hoft}
m_{T,\xi}(\theta )  \leq D T^{1- \ba(\tilde{\theta})} 
\qquad {\rm uniformly\; for}\; \theta\in {\mathcal I}_{\Delta} (\tilde \theta). 
\end{equation}

As we are dealing with an ergodic system, 
the bound (\ref{hoft}) holds also for $s_{T,\xi}(\theta ; z)$,  
for $\Omega$-a.e. $z\in {\mathcal M}_\vz$, because the dependence 
on $z$  in (\ref{sp-avera}) is actually missing.\footnote{More precisely, 
accordingly to \cite{hof}, the critical 
 exponent $1-\ba (\tilde{\theta})$ in (\ref{hoft}) is an upper bound for the exponent of $s_{T,\xi}(\theta ;
 z)$, for $\Omega$-a.e. $z\in {\mathcal M}_\vz$.}

 Consequently,  it follows from (\ref{dps}) that, for $\Omega$-a.e. $z$, 
 the sequence of functions  $\rho_T (\theta ;z)$ 
 as $T\to +\infty$ is bounded by:
\begin{equation}
\label{ass-t}
\rho_T (\theta ;z) \leq C T^{1-\frac {\ba (\tilde \theta) }{2}}\qquad 
{\rm for}\; \theta\in {\mathcal I}_{\Delta} (\tilde \theta),\;\; {\rm with}\;\; C>0; 
\end{equation}
in particular, for instance, if $\tilde \theta =0$ and $\Delta = 2{\pi}/{\bar T}$ (with $\bar T \geq 2$), 
(\ref{ass-t}) holds in every interval ${\mathcal I}_{\delta} (\theta)$  with $\delta \leq  2{\pi}/{T}$.

%
 %

\section{Analytic estimate for the moments' scaling function}
\label{result}

An accurate study of moments' asymptotic behaviour at large times  
is the central point of the present paper. 
The moments' scaling function $\gamma (n)$ is defined as the real number  $\gamma (n)$
such that the discrete Mellin transform:
\begin{equation}
\label{mellin}
I^{(n)} (\beta) = \sum _{t=1}^{+\infty} \frac {1}{t^{1+\beta}}\sigma ^{(n)} (t),
\end{equation}
converges for $\beta > \gamma (n)$ and diverges for $\beta < \gamma (n)$. 
This definition  of 
$\gamma(n)$, respect to other definitions based on the asymptotic behaviour 
of $\sigma ^{(n)} (t) $ as $t\to \infty$, has 
the advantage that it does not  take into 
account of possible subdominant 
contributions to the transport process.

The moments' scaling function may be written as  $\gamma (n) = \nu (n) n$. 
Normal transport and {\em weak} anomalous transport fall in 
the category with $\nu (n)=\nu_0=const.$ with $\nu_0 =1/2$ and $\nu_0\not=1/2$, respectively; 
at long times, the distribution of $\Xi(z,t)$ relaxes to a self-similar function, which is a Gaussian 
distribution when $\nu_0 =1/2$. 
{\em Strong} anomalous transport corresponds to the case where the distribution of 
$\Xi$ does not collapse to a self-similar form, multiple scales exist 
and a phase transition is observed, marked by a piecewise linear 
function $\nu (n)$ \cite{vulp,acri}. 

For even moments of integer order ( $n\geq 2$), we obtain an analytic relationship, linking 
the scaling function $\gamma (n)$ with H\"older exponent  at $\theta=0$ of the spectral measure 
associated to $\xi (z)$. 
This relation extends to higher order moments the formula found in \cite{agr} for 
  the exponent $\gamma(2)$ of the 2-nd moment.

By substituting (\ref{wk-n}) into (\ref{n-mo-1}), the moment of order $n$ is expressed by  
the following multiple integral over the $(n-1)$-dimensional torus $\TMN$: 
\begin{equation}
\label{moments-n}
\sigma ^{(n)}(t) =
 \frac {1}{(2\pi)^{n-1}} \int_{\TMN} \ddb \vt  \; m_\xi  (\vt  \;)
D_t \left( \Theta (\vt )\right) \prod _{j=1}^{n-1}
D_t (\theta _j), 
\end{equation}
where $D_t (x)$ denotes the kernel: 
\begin{equation}
\label{dirichlet}
D_t (x)= e^{-\frac i2 x(t-1)} \sum_{n=0}^{t-1}e^{inx}=
\frac{\sin\left(\frac {x}{2} t\right)}{\sin\left(\frac {x}{2} \right)}. 
\end{equation}

The conjungate symmetry property (\ref{mcc}) of the polyspectrum guarantees that 
the expression (\ref{moments-n}) is real.
Moreover, since $D_t(x)$ is an even function of $x$, the parity transformation property 
of the integrand in (\ref{moments-n}) is determined by $m_\xi (\vt )$. From 
 (\ref{m-odd-1}), it follows that all  moments of odd order are null. This is consistent 
with the fact that the angle $\Xi(z,t)$, accumulated by a single particle,  may assume positive or negative 
values;  hence, in \ref{up-odd}, odd order moments of $|\Xi (z,t) |$ are taken into account. 


 
%
%
For even $n$, owing to symmetry (\ref{m-even-1}),  
the integration domain of (\ref{moments-n}) can be reduced to half of the 
$(n-1)$-dimensional hyperoctans. 

By substituting (\ref{moments-n}) into  (\ref{mellin}), we get:
\begin{eqnarray}
\label{mellin-int}
\fl && I^{(n)}(\beta )=\frac {2}{(2\pi)^{n-1}} \sum _{l=1}^{2^{n-2}} I_l^{(n)}(\beta )\nonumber \\
\label{in}
\fl && I^{(n)}_l(\beta )=\int_{{\LM}_l^{n-1}}\ddb \; \vt 
 \; m_\xi  (\vt  ) S(\beta, \vt )
=\lim_{T \to \infty}   \int_{{\LM}_l^{n-1}}\ddb \; \vt 
 \; m_{T,\xi}  (\vt  ) S_T(\beta, \vt )  \\
\fl && S_T(\beta, \vt )= \sum _{t=1}^T \frac {1}{t^{1+\beta}}
D_t \left( \Theta (\vt )\right) \prod _{j=1}^{n-1}
D_t (\theta _j) \qquad \quad S(\beta, \vt )=\lim_{T \to \infty} S_T(\beta, \vt )\nonumber 
\end{eqnarray}
The integration domains $\LM _l^{n-1}$ in (\ref{in}) 
 are Cartesian products of $(n-1)$ half tori, $\TM _+ =[0,\pi]$ or 
$\TM _- =]-\pi,0]$. For instance for $n=4$,  $\LM _l^{3}$ are the following octants: 
$\LM _1^{3}=\TM _+ \times \TM _+ \times \TM _+ $, $\LM _2^{3}=\TM _+ \times \TM _+ \times \TM _- $, 
$\LM _3^{3}=\TM _+ \times \TM _-\times \TM _+ $ and $\LM _4^{3}=\TM _- \times \TM _+ \times \TM _+ $.

As shown below, the convergence of $I^{(n)}(\beta )$ is
guaranteed under the condition 
\begin{equation}
\label{cond-1}
\beta > n \left( 1 - \frac{\ba (0)}{2} \right); 
\end{equation}
$\ba (0)$ is the H\"older exponent  at 0 of the spectral measure, associated to $\xi$.

Therefore, according to the definition of the moments' scaling function $\gamma (n)$ defined 
via the discrete Mellin transform (\ref{mellin}), we have:
\begin{equation}
\label{result-n}
\gamma(n) = n\left( 1-\frac {\ba (0)}{2}\right) \qquad \qquad  n \; {\rm positive\; even\; integer}.
\end{equation}
As stated by (\ref{result-n}), the spectrum of moments is governed by a single scale, 
and no {\em strong}
anomalous transport takes place.

\subsection{Derivation of the convergence condition (\ref{cond-1})}

 The argument (for even order moments) consists of two different steps. 
The first step involves the second term appearing in (\ref{in}), namely $S_T(\beta,\vt)$, 
which constrains an evaluation of (\ref{in}) around the origin. Indeed,   
  $\lim _{t\to \infty }
D_t (x)=0$ uniformly for $| x |\geq \delta >0$ and the dominant contribution to the integral 
 (\ref{mellin-int}) comes from simultaneous zeros of the denominators of $D_t (\theta _j)$, 
 namely from $\theta _j=0$, $\forall j=1,\cdots n-1$ \cite{dana}.  


The kernel  $D_t(x)$ 
is an even function of $x$, whose smallest positive zero is given by $x=2\pi/t$; 
moreover, in the interval $|x |\leq 2\pi /t$: $0< D_t (x) \leq t$. It is sufficient to 
choose $|\vt | \leq 2\pi/((n-1)T)$ to get all the arguments of the kernels in $S_T(\beta,\vt)$ 
within the range $|x|\leq 2\pi /t$.  

 Around the origin,  we may thus write  
\begin{equation}
\label{sumttap}
S_T(\beta,\vt )\leq \sum_{t=1}^T \frac 1{t^{\beta+1}}t^{n} 
\equiv s(1,T), \qquad \quad {\rm for } \; \; |\vt | \leq \frac {2\pi}{(n-1)T}.
\end{equation}
 For $\beta <n$\footnote{For $\beta >n$ the integral (\ref{in}) is always 
 convergent in a small sphere around the origin.}, the partial sum is bounded by: 
\begin{equation}
\label{sum-ttt}
s(1,T)\leq c(\beta ) T^{n-\beta}, 
\end{equation}
where $c(\beta )$ is finite function for $\beta \neq n$. 

By using $T\leq 2\pi / ((n-1) |\vt |)$, the  final result for $S(\beta, \vt )$ is:
\begin{equation} 
\label{sum-t-n}
S(\beta, \vt ) \leq s(1,T) \leq  
c_1 (\beta ) |\vt | ^{\beta-n} 
\end{equation}
where: $c_1(\beta)=(2\pi /(n-1))^{n-\beta} c(\beta)$.  

Secondly, we 
%
derive the local scaling behaviour of  $m_\xi(\vt  )$ near a singularity $\vec {\tilde \theta}$
in frequency space  $\TM ^{n-1}$. As in 1-dimensional case, the 
scaling is related to the asymptotic growth of the sequence $\{ m_{T,\xi} (\vt )\}_{T\in \NMp}$ 
as $T\to \infty$. This link holds locally in frequency space, inside a small 
$(n-1)$-sphere centered at the singularity of radius $|\delta \vt  |= |\vt - \vec {\tilde \theta}|
\leq 2\pi/(T(n-1))$; 
then for large times, i.e. $T\to \infty$, $|\delta \vt |\to 0$. 

The high-dimensional spectrum (\ref{mtn}) is 
expressed by a phase average of the contributions of different trajectories, i.e. $M_{T,\xi}
(\vt ; z_{s,l})$.  Formula
(\ref{m-even}) gives 
$M_{T,\xi} $ as a function of $\rho_T$ and 
$\phi_T$ of a single variable $\theta$, where $\theta$ means $\theta _j$ with $j=1,\cdots , n-1$ 
or $\Theta$.  By  making use of the asymptotic behaviour 
(\ref{ass-t}) of the modulus $\rho_T $, valid  for $T\to \infty$, 
 from (\ref{m-even}) we get the bound: 
\begin{equation}
\label{asy-n-time}
 \fl | M_{T,\xi} ({\theta}_1, \cdots , {\theta}_{n-1} ; z_{s,l}) |\leq C'  T ^{n-1} 
T^{-\ba (\tilde{\Theta})/2}\prod_{i=1}^{n-1}T^{-\ba(\tilde{\theta}_i)/2},
\end{equation}
and, in particular, in a small $(n-1)$-sphere centered at $\vec 0$, 
\begin{equation}
\label{asy-n-theta}
  m_\xi(\vt )  \leq C 
|\vt  |^{n\left( \frac {\ba(0)}{2}-1\right) +1} \qquad\quad  {\rm for} \;\; 
 |\vt  |\leq \frac {2\pi}{(n-1)T}.
\end{equation}
In lhs of (\ref{asy-n-theta}) we omit the absolute value because, 
as explained in section \ref{link}, for a fixed $T$, the modulus $\rho_T$ 
and the phase $\phi_T$ are continuous functions of $\theta$ and in particular, since 
$\phi_T(\theta=0)=0$, 
$\phi_T\to 0$ for $\theta \to 0$. Hence the scaling behaviour  in $\vec 0$ 
 of the total phase of $M_{T,\xi}$, which is 
a sum of the single phases,  is a trivial. 
%
%

We finally derive condition (\ref{cond-1}), under which the integrals (\ref{in}) 
converge. 

 We consider a $(n-1)$-sphere of radius $2\pi /((n-1)T)$, centered in the 
origin of the frequency space, inside which the integrand function is bounded by 
(\ref{sum-t-n}) and  (\ref{asy-n-theta}). Dominant contributions to  integrals (\ref{in}) 
are restricted  inside this $(n-1)$-dimensional sphere. 

The  integrals may be
 evaluated by $(n-1)$-dimensional hyperpherical 
coordinates $\{ r, \psi _1, \psi _2,\cdots \psi_{n-2}\} $ \cite{polar}. 
 The 
vector $\vt $ may be written as $\vt =r\vec \omega$; $0\leq r\leq 2\pi /((n-1)T)$ and 
 $\vec \omega$ is a unity vector with components: 
$\omega_1 =\cos \psi_1, \omega_2 = \sin\psi_1 \cos\psi_2; \; 
\cdots ; \; \omega_{n-2}=\sin \psi_1 \sin\psi_2 \cdots \sin \psi_{n-3}
\cos\psi_{n-2}; \;  \omega_{n-1} =\sin\psi_1\sin\psi_2\cdots \sin\psi_{n-2}$ with 
 $0\leq \psi_l\leq \pi$ for $l=1,n-3$ and $0 \leq \psi_{n-2} \leq 2\pi$.  
The 
Jacobian of the coordinate transformation is $J=r^{n-2}\sin ^{n-3}\psi_1 
\sin ^{n-4}\psi _2 \cdots \sin \psi_{n-3}$.

 The  integration over variables 
$r$ and $\vec\omega$ can be carried on independently. From  (\ref{sum-t-n}) 
and  (\ref{asy-n-theta}), we may write:
\begin{equation}
\label{split-int}
\fl \left. I^{(n)}_l (\beta ) \right\vert _{|\vt | \leq \frac {2\pi}{(n-1)T}} \leq  c_2(\beta )
\int _{s_{n-2}} \dd \vec \omega  
 \int _{0}^{\frac {2\pi}{(n-1)T}} \dd r \; r ^{\beta -n\left( 1-\frac {\ba (0)}{2} \right)-1}
\end{equation}
 $s_{n-2}$ is a $(n-2)$-dimensional hypersphere of unitary radius and $c_2(\beta )$ 
 is a finite function for $\beta \neq n$.

The integral over the angles is trivial, while, the radial part
 %
%
is convergent under the condition (\ref{cond-1}): $\beta > n(1-\ba (0)/2)$, which
finally yields the estimate for the exponent of even order moments.



\begin{figure}
 \begin{center}
  \includegraphics[width=8cm,angle=0]{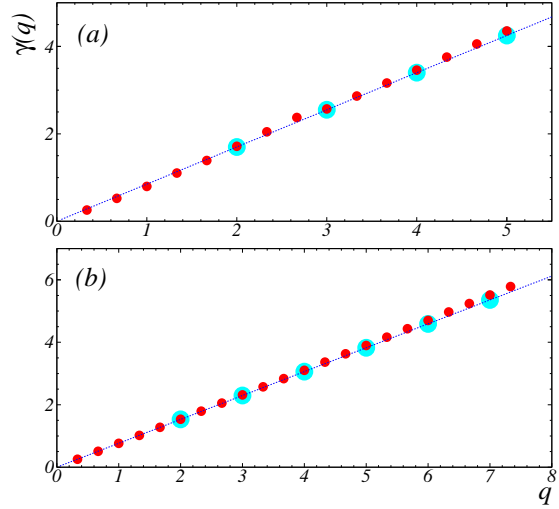}
  \caption{(Colour online) Scaling function $\gamma (q)$ of the moments 
  $ \sigma ^{(q)}(t)$, given by (\ref{MMq}),  plotted vs the order $q$. 
  Dotted straight lines 
  have a slope $q\left( 1-\frac {\ba (0)}{2} \right)$, according to the theoretical prediction (\ref{result-n}), 
  extended to real values of $q$. Moments of integer orders are marked by halos. 
  (a) refers to same data of previous figures, for which  the scaling exponent $\ba (0)= 0.3$
   is derived in 
  figure \ref{ms-alp}. The parameters of (b) are: $l_1/l_2 = (\sqrt 5+1)/2$, $\tan\varphi=\pi/4$ and 
  $\ba (0) =0.47$.}
\label{moh}
 \end{center}
\end{figure}

\section{Numerical simulations and conclusions}
\label{num}

 From the numerical point of view, we may extend the analysis to moments of 
 the diffusive variable $|\Xi (z,t)|$ 
of arbitrary positive real order. The $q$-th order moment is 
\begin{equation}
\label{MMq}
\sigma^{(q)}(t) \,=\, \int _{{\mathcal M}_\vz} \dd \Omega (z)
 \left| \Xi(z,t)\right|^q , \qquad q\in {\RM}_+ ; 
\end{equation}
the scaling function $\gamma (q)$ is defined by (\ref{mellin}), with $\sigma ^{(n)}$ 
replaced by $\sigma ^{(q)}$.
Note that in spite of having introduced the absolute value respect to 
the definition (\ref{n-mo}), we keep the same notation. 

Numerical simulations, presented throughout the paper,
  refer to a billiard table belonging to {\it class 3} with parameters $l_1/l_2 = \pi /2$ 
and $v_x/v_y = \tan \varphi =\pi /4$; a second case is shown in figure  \ref{moh} (b), referring to 
$l_1/l_2 = (\sqrt 5 +1)/2$. 

The analytical expression for the exponent $\gamma (n)$ of the algebraic growth 
 of integer order moments, given by (\ref{result-n}), is numerically tested in figure \ref{mo4}, 
 in which the $4$-$th$ order moment is plotted as a function of time $t$ in logarithmic scales. 
 The straight line has a slope $\gamma (4)=3.4$, given by the formula 
 with $n=4$ and $\ba (0)=0.3$.  Numerical procedure to derive $\ba (0)$ is explained below.

Figure \ref{moh} displays the scaling function $\gamma (q)$ of absolute moments of arbitrary real order 
 (\ref{MMq}), for the 
 two parameters'  pairs $(l_1/l_2 , \tan \varphi )$ specified above. 
  Numerically evaluated exponents are shown by dot symbols as
  a function of $q$; they are 
    calculated by a linear least-square fit of $\ln \sigma ^{(q)} (t)$ versus $\ln t$.  
    
  Analytical estimates of the exponents of  integer order moments 
  are shown by halos, surrounding the dots. 
   Inside the inspected range,  numerical data are consistent with 
a linear dependence of $\gamma (q)$ on $q$. The ratio $\gamma (q)/q$ appears 
to be constant for fixed parameters $(l_1/l_2 , \tan \varphi )$ and to depend only on the 
spectral index $\ba$ in $\theta =0$. This ratio ranges between $1/2$ (normal diffusion) 
and $1$ (ballistic dispersion). Therefore the square annulus billiard displays a 
so called ``weak" anomalous diffusive (or  strong self-similar) process \cite{vulp,fmy01}.  

We finally provide a few further details on the numerical procedures. 
 
The discrete-time directional dynamics at fixed $\varphi$ is evolved up to a time $T=2^{20}$ 
and the number of initial conditions 
employed in phase averages ranges from $10^4$ to $10^6$.

An approximation of the microcanonical measure inside the phase space 
${\mathcal M}_\varphi$ is obtained as follows. The phase average is evaluated by
taking a  uniform distribution of the particles inside 
the accessible region of the billiard and the sign of the components of the 
unity velocity vector $\vec v$  is assigned 
at random. 
 The first collision points with the boundaries of the billiard are taken as 
initial conditions of the Birkhoff mapping.  

 Spectral analysis of the signals is reconstructed numerically by employing the 
Fast Fourier Transform algorithm (FFT), i.e. a finite discrete Fourier Transform for vectors in 
$\CM ^{2T}$. 
Two different methods  have been tested to calculate the coarse-grained approximation of the 
(1-dimensional) average density power spectrum (\ref{sp-avera}).  According to 
(\ref{wk-n}), $m_\xi (\theta)$ may be approximated by a direct-FFT 
 of the phase-averaged correlation sequence $C^{ph}_\xi (t)$. 
  Alternatively, according (\ref{sp-avera}) and (\ref{dps}), we calculate the square modulus 
  of the direct-FFT of the signal $\xi (B^t z)$ of single trajectories and then make a phase 
  average over the initial conditions. Owing to the finite time interval $-T\leq t\leq T$, 
  we multiply time sequences, $C^{ph}_\xi (t)$ or $\xi (B^t z)$, 
  by a proper windowing function \cite{har}; 
  to reduce the amount of  negative values in the former method, we prefer to use 
   a triangular window 
function $w_t =1-|t|/T$,  instead of a square 
windowing function as  in the theoretical treatment, because its partial Fourier series is
 a non negative function. 
  In both methods, the resolution in frequency space is $\Delta \theta =\pi/T$. 
 For the finest resolution, i.e. $T=2^{20}$, the order of magnitude of negative values, in the first method,
 is $\lesssim 10^{-4}$. Apart from 
numerical errors, the two methods give the same results, for a fixed resolution. 
By comparison of the two methods, we get that less than the $1\%$ of the total mass on the 
torus is affected by numerical errors. 
In figure  \ref{ms-alp} (a) a reconstruction of $m_\xi (\theta)$ is shown, calculated with a 
resolution of $\pi/(2^{14})$.
 
The first two generalized dimensions $D_1$ 
(information dimension)  and 
$D_2$ (correlation dimension) \cite{proca}  are calculated 
 by making a sequence of dyadic partitions of the coarse-grained approximation 
of $m_\xi (\theta )$.
At step $N$, the interval $(-\pi,\pi]$ is divided into $2^N$ sub-intervals 
$I_{N,j}$ ($j=1,\cdots , 2^N$) of length $\delta_N =2\pi/2^N$, each of which contains a mass 
$\mu (I_{N,j})$. As $N\to \infty$ (i.e. $\delta _N\to 0$), $D_1$ and $D_2$ 
 are defined by:
\begin{eqnarray}
\label{def-d1d2}
&&  \chi _1 (N)= {\sum_{j=1}^{2^N} \mu (I_{N,j}) \ln \mu (I_{N,j}) } \sim  D_1\; {\ln \delta_N}
\nonumber \\
&&  \chi _2 (N) = {\ln \sum_{j=1}^{2^N} \mu ^2 (I_{N,j})} \sim D_2 \;  {\ln \delta_N}. \nonumber 
\end{eqnarray}
We extrapolate $D_1$ and $D_2$ by a linear fitting in logarithmic scale  with $N$  ranging from 
$N=2$ to $N=20$; an example is shown in figure \ref{co-d1d2} (b). 
A value of $D_1 <1$ is found,  consistent with the presence of a 
singular continuous component of the spectrum. As shown in figure \ref{co-d1d2} (a), 
the obtained value of $D_2$ gives a satisfactory estimate of the power law decay exponent 
of the integrated correlation function. 

The scaling index $\ba$ of the spectral measure in $\theta =0$ 
is also evaluated numerically. 
 In figure \ref{ms-alp} (b) the mass contained in the interval $[-\delta , \delta]$ is shown 
 as a function of $\delta$. Intervals of different widths $\delta _M =M\Delta \theta$ are considered, 
 with a resolution  $\Delta \theta$ ranging from $\pi/ 2^{20}$ to $\pi/ 2^6$; different symbols refer to 
 $M=0$ (triangles), $M=1$ (squares) and $M=8$ (circles). The mass is 
given by $\mu ([-\delta_M , \delta_M])=\sum _{j=-M}^M m _\xi (j\Delta \theta)$ and 
it is supposed to scales as $\sim \delta_M ^{\ba (0)}$. 
  The straight line, which fits the data in logarithmic scales in the range $-11 < ln \delta < -7$,  
 has a slope of $\ba (0)=0.30\pm 0.05$. The data in the left part of figure \ref{ms-alp} (b) are 
 not taken into account in the derivation of $\ba (0)$ because the corresponding mass is 
 close or below the numerical precision of the simulations.  The same calculation 
 in case $l_1/l_2 = (\sqrt 5+1)/2$ gives the value $\ba (0)= 0.47\pm 0.05$. Note that the error 
 on the numerical estimate of $\bar \alpha (0)$ is not relevant when compared with data of 
 figure \ref{moh}.

We have thus provided analytic arguments and numerical support that polygonal billiards
 may exhibit anomalous transport in a weak sense \cite{vulp}:  the distribution of the angle, accumulated by single particles at fixed time, is not gaussian and  the moments' asymptotic growth is governed by a single scale, which is related to the H\"older exponent of the spectral measure at zero. 
 
This work has been partially supported by MIUR-PRIN project ``Nonlinearity and 
disorder in classical and quantum transport processes".  
  

\section*{Appendices}

\appendix

\section{Power spectrum, bispectrum and trispectrum}

\label{cor-pol}

For $n=2$, namely for  a single frequency $\theta\in \TM$, the result (\ref{dps}) 
 is retrieved from (\ref{m-even}); the weak limit $s_\xi (\theta; z)$ of the sequence (\ref{dps}) 
 is the density power spectrum of the signal.
 
For the first few orders, i.e. $n=3$ and $4$ (bispectrum and trispectrum),  
we have the following explicit expressions, respectively:
%
%
\begin{eqnarray}
\label{m-3}
\fl && M_{T,\xi} (\theta_1,\theta_2; \elll )=\frac {i}{2T} 
\rho_T(\theta_1;\elll ) \rho_T(\theta_2;\elll )\rho_T(\theta_1+\theta_2;\elll ) \cdot 
\nonumber \\
\fl  && \qquad\qquad\qquad\qquad
 \cdot \sin( \phi_T(\theta_1; \elll )+
\phi_T(\theta_2; \elll ) -  \phi_T(\theta_1+\theta_2; \elll ) ),
\end{eqnarray}
and
%
%
%
\begin{eqnarray}
\label{m-4}
\fl M_{T,\xi}(\theta_1,\theta_2,\theta_3; \elll )&=&\frac {1}{2T} 
\rho_T(\theta_1; \elll ) \rho_T(\theta_2; \elll ) \rho_T(\theta_3; \elll ) 
\rho_T(\theta_1+\theta_2+\theta_3 ; \elll ) \cdot \nonumber \\ 
&&\quad
\fl  \cdot \cos( \phi_T(\theta_1; \elll )+
\phi_T(\theta_2 ; \elll )+\phi_T(\theta_3 ; \elll ) - \phi_T(\theta_1+\theta_2+\theta_3; \elll ) ), 
\end{eqnarray}   
with $\elll =(s,\varphi_l)$ and $\varphi_l =\varphi -l\pi /2$ ($l=0,1$).

\subsection{Symmetry properties}
\label{sim}

Owing to the symmetry properties of higher order spectra and of multiple-time
 phase averaged correlation functions (\ref{cphasen}), 
 the integration domain in (\ref{wk-n}) can be reduced. Some properties, as (\ref{mcc}), 
(\ref{m-odd-1}) and (\ref{m-even-1}),  follow directly from (\ref{poly-n-ap}) and 
(\ref{even}), (\ref{odd}) and have been used to restrict the domain in (\ref{in}), as explained 
in section \ref{result}.

A general procedure to determine the principal domain of polyspectra, 
i.e. the nonredundant region of computation, is explained in \cite{pdom3,pdom,pdom2}.
The symmetry properties of higher order spectra can be grouped into:
\begin{enumerate} 
\item {invariance under permutation of any pair of frequencies;}
\item {the conjugate symmetry property (\ref{mcc}), which implies the redundancy of half of 
some frequency axes;}
\item {periodicity of period $2\pi $ is each frequency variable; this is a consequence of the 
fact that the partial sums (\ref{fou}) are periodic functions of the frequencies 
 $ \theta_i \in \TM$ ($i=1,\cdots ,n$)}.
\end{enumerate}
In this appendix, we  review some symmetry conditions for the 2- and 3- dimensional cases \cite{sim1}.

\noindent  {\it{ Two dimensional case.}}   
From the definition of 2-time phase-averaged correlation function, i.e.  (\ref{cphasen}) with $n=3$, 
and the invariance of the measure $\dd \Omega (z)$
under $B^t_\varphi$, we get:
\begin{equation}
\label{c2-sim}
\fl C^{ph}(t_1,t_2)=C^{ph}(t_2,t_1)=C^{ph}(-t_1,t_2 -t_1)=
C^{ph}(-t_2,t_1-t_2).
\end{equation}
From  (\ref{m-3}), (\ref{even}) 
and (\ref{odd}), we have:
\begin{equation}
\label{sim-n2}
\fl m (\theta_1,\theta_2)=m  (\theta_2,\theta_1)=m^*(-\theta_1,-\theta_2)=m (\theta_1,-\theta_1 -\theta_2)=
m (\theta_2,-\theta_1 -\theta_2).
\end{equation}
Owing to (\ref{sim-n2}), the bispectrum is symmetric about the lines $\theta_1=\theta_2$, 
$\theta_1+2\theta_2=2\pi n_1$ and $\theta_2+2\theta_1=2\pi n_2$ ($n_1,n_2 \in \ZM$). 
The principal domain is the triangle of vertices $A(0,0)$, $B(\pi ,0)$ and $C(\frac 23 \pi, \frac 23\pi)$, 
i.e. $0\leq \theta_1\leq \frac 23 \pi ,\; 0\leq \theta_2 \leq \theta_1$ and 
$\frac 23 \pi \leq \theta_1\leq \pi ,\; 0\leq \theta_2 \leq -2\theta_1+2\pi$.

\noindent  {\it {Three dimensional case.}}   
The 3-time phase-averaged correlation function satisfies:
\begin{equation}
\label{c3-sim}
\fl C^{ph}(t_1,t_2,t_3)=C^{ph}(t_1,t_3,t_2)=
C^{ph}(-t_1,t_2-t_1,t_3-t_1); 
\end{equation}
and the trispectrum:
\begin{equation}
\label{m3-sim}
\fl m(\theta_1,\theta_2, \theta_3)=m(\theta_1,\theta_3,\theta_2)=
m (-\theta_1,-\theta_2, -\theta_3)=m(\theta_2, \theta_3,
-\theta_1 -\theta_2-\theta_3).
\end{equation}
Cyclic relations, obtained by permutations of $(t_1,t_2,t_3)$ in (\ref{c3-sim}) 
and $(\theta_1,\theta_2,\theta_3)$ in (\ref{m3-sim}), hold. The trispectrum is symmetric 
about the plane $2\theta_1+\theta_2+\theta_3=2\pi n_1$ (and cyclic).  
The principal domain is a polyhedron with vertices $A(0,0,0)$, $B(\frac \pi2,\frac \pi2,\frac \pi2)$, 
$C(\pi,0,0)$, $D(\frac 23\pi, \frac 23\pi,0)$, $E(\pi,0,-\frac \pi2)$ and $F(\pi,\pi,-\pi)$ \cite{pdom}.


\section{Dynamical and spectral exponents.}
\label{kernel-hof}

 In this appendix we sketch how to derive (\ref{hoft}). 

We assume that the spectral measure $\dd\mu _\xi (\theta)$, associated to $\xi$, is 
 uniformly $\ba$-H\"older continuous (U$\ba$H)  in an interval 
${\mathcal I}_\Delta (\tilde \theta)\equiv [\tilde\theta -\Delta , \tilde\theta +\Delta]$, 
namely  a positive constant $c$ exists  such that 
 \begin{equation}
 \label{uah-con}
 \mu _\xi ({\mathcal I}_\delta (\theta) )\equiv 
 \int _{\theta -\delta }^{\theta +\delta } \dd \mu _\xi (\theta ' )\leq c\delta ^{\ba (\tilde \theta)}, 
 \qquad\; \; \forall {\mathcal I}_\delta (\theta)\subset {\mathcal I}_\Delta (\tilde \theta). 
\end{equation}

The sequence $\{ m_{T,\xi} (\theta )\}_{T\in \NMp}$ is defined by (\ref{sp-avera}) and (\ref{dps}) as
\begin{eqnarray}
\label{mt2-bis}
m_{T,\xi} (\theta ) &=&\int _{{\mathcal M}_\varphi} \dd \Omega (z) s_{T,\xi} (\theta ;z) \\
&=& \frac 1{2T} \int _{{\mathcal M}_\varphi} 
\dd  \Omega (z) \hat \xi_T(\theta ;z) \hat \xi^*_T(\theta ;z). \nonumber 
\end{eqnarray}
By making use of (\ref{fou}) and then of (\ref{cphase}), (\ref{wk-n}) and (\ref{ps-ms}), we obtain 
\begin{equation}
\label{m-ker}
\fl m_{T,\xi} (\theta ) = 
\frac 1{2T} \sum_{t =-T}^{T-1}  \sum_{t' =-T}^{T-1} 
C^{ph}_\xi (t-t') e^{-i\theta (t -t')}=
\frac 1{2\pi} \int _{\TM} \dd \mu _\xi (\theta ') K_{T} (\theta -\theta ')
\end{equation}
with 
\begin{eqnarray}
\label{fejer}
\fl K_{T} (x) &=& \frac 1{2T} \sum_{t_1=-T}^{T-1}  \sum_{t_2=-T}^{T-1} e^{ix(t_1-t_2)}
= \frac {1}{2T} \left( d_T^2 (x)-2d_T (x) \cos Tx+1 \right) =\nonumber \\
\fl &=& \frac 1{2T} \left(  \frac {\sin Tx}{\sin \frac x2 } \right)^2;
\end{eqnarray}
$K_{T} (x)$ is the Fej$\acute{\rm e}$r kernel, up to a constant factor, and  
$d_T(x)$ is the Dirichlet Kernel:
\begin{equation}
\label{diri}
d_{T} (x)  = \sum_{t=-T}^{T}  e^{ixt}
= \frac {\sin\left[ (2T+1)\frac x2\right]}{\sin\frac x2}.
\end{equation}

The kernel (\ref{fejer}) is a positive, periodic  even function of $x\in \TM $; moreover, it fulfills: 
\begin{eqnarray*} 
&(i)& \; K_{T} (0)=2T;  \\
&(ii)& \; \lim _{T\to\infty} K_{T} (x)=0 
\; \; {\rm uniformly} \; \; \; |x|\geq \delta >0;  \\
&(iii)& \; 0\leq K_{T} (x)\leq 2T\;\;\; \;\;\; |x|\leq \frac \pi T;  \\
&(iv)& \;   0\leq K_{T} (x)\leq  \frac {T}{2 k^2 }\;\;\; \;\;\; k\frac \pi T \leq |x|\leq (k+1)\frac \pi T
\;\;\; {\rm and}\;\; k\geq 1.  \\
 &(v)& \;  \frac {8T}{\pi ^2}\leq K_{T} (x) \leq 2T \;\;\; \;\;\; |x|\leq \frac {\pi}{2T}.
\end{eqnarray*} 

By using the properties above, the proof of 
 theorem 2.1 in \cite{hof} can be reproduced  for $m_{T,\xi }(\theta)$.  
 Note that $m_{T,\xi} (\theta )$ corresponds to $G_T(\theta)$, and $\beta$ 
 to $1-\ba$ in \cite{hof}. 
 
 Let $\theta \in {\mathcal I}_\Delta (\tilde \theta)$ and $0<\delta <\pi$, 
 we show that (\ref{hoft}) holds in an arbitrary interval 
 ${\mathcal I}_\delta ( \theta )\subset {\mathcal I}_
 \Delta (\tilde \theta)$.   ${\mathcal I}_\delta ( \theta  )$ 
 can be covered by a finite sequence of intervals: (1) 
 $ {\mathcal I}^{(0)} \equiv \{ \theta ' \in \TM / \; | \theta -\theta '  |\leq \frac \pi T\}$, in which $(iii)$ holds, and 
 (2) ${\mathcal I}^{(k)} \equiv \{  \theta ' \in \TM  /\;  k\frac  \pi T \leq\ | 
 \theta - \theta ' | \leq (k+1)\frac \pi T\leq \delta \}$, in which $(iv)$ holds. 
 
 Moreover,  assumption (\ref{uah-con})
 implies that the mass in every interval of length $\pi /T$, included in ${\mathcal I}_\Delta (\tilde \theta)$, 
 is bounded by $\mu _\xi ({\mathcal I}_{\frac {\pi}{2T}} (\theta))\leq D T ^{-\ba (\tilde \theta)}$, 
 with constant $D>0$ and $T>\bar T(\Delta, \theta)$; $\bar T$ is chosen so that for $T>\bar T$: 
 ${\mathcal I}_{\frac {\pi}{2T}} ( \theta )\subset {\mathcal I}_
 \Delta (\tilde \theta)$. 
 
 Therefore, 
 \begin{eqnarray}
 \label{hof-dimo}
\fl m_{T,\xi} (\theta ) \left |_{{\mathcal I}_\delta ( \theta)}\right. &=&
 \frac {1}{2\pi} \left\{ \int _{{\mathcal I}^{(0)}} +\sum _k 
 \int _ {{\mathcal I}^{(k)}  }  \right\}
 \dd \mu _\xi (\theta ') K_{T} (\theta -\theta ') \nonumber \\
 &\leq& \left( C_1  + C_2 
 \sum _{k=1}^{[\delta T/\pi]}\frac {1}{k^2}\right)  
 T ^{-\ba (\tilde \theta)} T  \leq C T ^{1-\ba (\tilde \theta)};
  \end{eqnarray}
$[.]$ denotes the integer part.


\section{Upper bound for absolute moments of odd order}
\label{up-odd}

As mentioned in section \ref{result},  the moments of odd order of the 
observable $\Xi (z,t)$ 
 vanish due to the symmetry property (\ref{m-odd-1}) of $m_\xi (\vt)$; we therefore 
 consider integer moments 
of the absolute value of the observable $\Xi (z,t)$:
 \begin{equation}
  \label{mo-n-abs}
   \sigma ^{(n)}(t)=\int _{{\mathcal M}_\varphi}\dd  \Omega (z) \left| 
  \Xi (z,t )\right|^{n}.
  \end{equation}
For even $n$ (\ref{mo-n-abs}) reduces to (\ref{n-mo}); we are interested to $n$ odd integer 
($n>1$). 

Since $\left| \Xi (z,t) \right| \leq \sum_{s=0}^{t-1}\left| \xi (B_\vz ^s ;z) \right|$, 
the following bound holds:
 \begin{equation}
 \label{majo}
   \sigma ^{(n)}(t)\leq \bar \sigma ^{(n)}(t)=
\int _{{\mathcal M}_\varphi} \dd \Omega (z) \left( \sum_{s=0}^{t-1}\left| \xi (B^s z) \right|\right)^{n}.
 \end{equation}
Therefore, the exponent $\bar \gamma (n)$ of the algebraic growth of $\bar\sigma^{(n)}(t)$ 
 constitutes an upper bound for the 
 exponent  $\gamma (n)$ of (\ref{mo-n-abs}); $\bar \gamma (n)$ 
 is again defined via a discrete Mellin transform (\ref{mellin}), 
 with $\sigma$ replaced by $\bar\sigma$. 
  
  We reproduce calculations in section  \ref{hstat}, by starting from the new 
  phase observable $\bar \xi  (B^t z)\equiv \left| \xi (B^t z)\right|$. The partial sum
  of Fourier series, i.e.  $ \hat {\bar \xi} _T (\theta; z)=\sum _{t=-T}^{T-1} |\xi (B^t z)| e^{-i\theta t}$, 
   verifies the property (cf. (\ref{fou-ti})): 
  \begin{equation}
  \label{fou-ti-disp}
  \hat {\bar \xi} _T (\theta; (s,-\varphi))= + e^{i\theta }\hat {\bar \xi }^* _T (\theta;(s,\varphi)). 
    \end{equation}
 By using (\ref{fou-ti-disp}) in the phase average, 
  the following expression for $\bar M_{T,\xi}$ is found (cf. (\ref{m-odd}) and (\ref{m-even})):
 \begin{equation}
\label{m-odd-abs}
\fl \bar M_{T,\xi} (\vt  ; \elll  )=\frac {1}{2T}
\bar \Gamma _T (\vt ; \elll )\bar \rho_T (\Theta (\vt); \elll )\cos \left( \bar \Phi _T ( \vt ; \elll ) - 
 \bar \phi_T ( \Theta (\vt ) ; \elll ) \right).
\end{equation}
 $\bar M_{T,\xi}$ is now an even function of $\vt $ and it shares the same symmetry properties 
of (\ref{m-even}).

For $n=3$, in particular we have:
\begin{eqnarray}
\label{m-3-abs}
\fl && \bar M_{T,\xi} (\theta_1,\theta_2; \elll )=\frac {1}{2T}
\bar \rho_T(\theta_1; \elll ) \bar\rho_T(\theta_2; \elll )\bar \rho_T(\theta_1+\theta_2; \elll ) 
\cdot \nonumber \\
\fl && \qquad\qquad\qquad\qquad
\cdot \cos(  \bar\phi_T(\theta_1; \elll )+
\bar\phi_T(\theta_2; \elll ) - \bar\phi_T(\theta_1+\theta_2; \elll ) ).
\end{eqnarray}

 The exponent $\bar \gamma (n)$ can be derived by proceeding as in section \ref{result}; 
 by denoting $\bar {\bar \alpha}(0)$ the H\"older index of $\dd \mu_{|\xi |}$ 
in 0, we get 
 $\bar \gamma (n)=n(1-\bar {\bar \alpha}(0)/2)$.

The mass associated to $|\xi|$ in a small interval, centered in 0, is 
\begin{equation}
\label{mass-abs}
\fl \mu_{|\xi |}({\mathcal I}_\delta (0)) =\lim _{T\to \infty} \frac {\delta}{T} \int_{{\mathcal M}_\varphi} 
\dd \Omega (z) 
\sum _{t=-T}^{T-1} \sum _{t'=-T}^{T-1} |\xi (B_\varphi ^t z) \xi (B_\varphi ^{t'} z) | 
\; {\rm sinc} (\delta (t-t')), 
\end{equation}
with ${\rm sinc} (x) =\sin x/x$.  If $\delta \leq \pi / (2T)$, $\mu_{\xi }({\mathcal I}_\delta (0)) 
\leq \mu_{|\xi |}({\mathcal I}_\delta (0))$ and therefore $\bar {\ba } (0)\leq \ba (0)$.




\end{document}